\documentclass[traditabstract]{aa}
\usepackage{amssymb,amsmath,epsfig,bm,color,slashed,mathrsfs}
\usepackage{amsmath,bm}
\usepackage{txfonts}
\usepackage{natbib}
\usepackage{graphicx}
\usepackage{multirow}
\usepackage{slashed}
\bibpunct{(}{)}{;}{a}{}{,}
\newcommand{\be}{\begin{equation}}
\newcommand{\ee}{\end{equation}}
\newcommand{\bea}{\begin{eqnarray}}
\newcommand{\eea}{\end{eqnarray}}

\newcommand{\tauvec}{{\bm \tau}}
\newcommand{\rhovec}{{\bm \rho}}

\newcommand{\eg}{{\it e.g.}}

\definecolor{red}{rgb}{0.8,0,0}
\definecolor{violet}{rgb}{0.4,0,0.4}
\definecolor{green}{rgb}{0,0.5,0.0}
\definecolor{navy}{rgb}{0.0,0.0,0.6}
\definecolor{orange}{rgb}{0.8,0.2,0.0}
\definecolor{blue}{rgb}{0.3,0.0,0.8}

\begin{document}
\title{Rotating  hybrid compact stars}

\author{N. S. Ayvazyan \inst{\ref{YSU},\ref{JWG}}
\and
G. Colucci \inst{\ref{JWG}}
\and 
D. H. Rischke \inst{\ref{JWG},\ref{FIAS}}
\and
 A.  Sedrakian \inst{\ref{JWG},\ref{YSU},\ref{AEI}}
}
\institute{Department of Physics, Yerevan State University, Alex
Manoogian 1,  0025  Yerevan, Armenia \label{YSU}
\and
 Institute for Theoretical Physics,
 J. W. Goethe University, D-60438 Frankfurt am Main, Germany
\label{JWG}
\and
Frankfurt Institute for Advanced Studies,
 J. W. Goethe University, D-60438 Frankfurt am Main, Germany
\label{FIAS}
\and
Max-Planck-Institut f\"ur Gravitationsphysik, Albert-Einstein-Institut, Potsdam, D-14476, Germany\label{AEI}
}

\titlerunning{Rotating hybrid stars}
\authorrunning{N. S. Ayvazyan et al.}

%\date{\textcolor{red}{\bf DRAFT: \today}}

\abstract{ Starting from equations of state of nucleonic and
  color-superconducting quark matter and assuming a first-order phase
  transition between these, we construct an equation of state of
  stellar matter, which contains three phases: a nucleonic phase, as
  well as two-flavor and three-flavor color-superconducting phases of
  quarks. Static sequences of the corresponding hybrid stars include
  massive members with masses of $\sim 2 M_{\odot}$ and radii in the
  range of $13\le R\le 16$ km. We investigate the integral parameters of
  rapidly rotating stars and obtain evolutionary sequences that
  correspond to constant rest-mass stars spinning down by
  electromagnetic and gravitational radiation. Physically new
  transitional sequences are revealed that are distinguished by a
  phase transition from nucleonic to color-superconducting matter for
  some configurations that are located between the static and
  Keplerian limits. The snapshots of internal structure of the
  star, displaying the growth or shrinkage of superconducting volume
  as the star's spin changes, are displayed for constant rest mass
  stars. We further obtain evolutionary sequences of rotating
  supramassive compact stars and construct pre-collapse models that
  can be used as initial data to simulate a collapse of
  color-superconducting hybrid stars to a black hole.  }
% 26.60.+c      Nuclear matter aspects of neutron stars in nuclear physics
% 97.60.Jd      Neutron stars
% 21.65.+f  Nuclear matter
%13.15.+g Neutrinos: interactions, 13.15.t +g
%\pacs{97.60.Jd,26.60.+c,21.65.+f,13.15.+g}


\maketitle

\section{Introduction}

%---------------- General motivation -------------------

It has been suggested long ago that compact stars could be dense
enough for a transition from baryonic matter to deconfined
quarks~\citep{1965Ap......1..251I,1970PThPh..44..291I,1974NuPhA.219..612I,1975PhRvL..34.1353C}.
If so, the properties of matter should be described in terms of
colored quarks as the fundamental degrees of freedom.  Therefore,
compact-star phenomenology offers a unique tool to address the
challenge of understanding the phase structure of dense quantum
chromodynamics (QCD).  The integral parameters of compact stars, such
as the mass, radius, and moment of inertia, depend sensitively on
their theoretically deduced equation of state (hereafter EoS) at high
densities. Currently, the measurements of pulsar masses in binaries
provide the most clean and stringent observational constraints on the
underlying EoS.  Important examples are the mass measurements
of two solar mass pulsars, PSR J1614-2230 and PSR J0348+0432, using
independent methods~\citep{2010Natur.467.1081D,2013Sci...340..448A}.

Because of their compactness neutron stars can rotate extremely
fast. Among them, millisecond pulsars offer a unique laboratory for
simultaneous measurements of high spins, masses, and moments of
inertia in binary systems. The last two integral parameters depend on
the EoS  of dense matter and thus shed light on its
properties. In particular, the Keplerian limit at which the star
starts shedding mass from the equator depends on the EoS.  Thus, a way
to constrain the EoS of dense matter is provided by observations of
high-spin pulsars.  The Keplerian frequency sets an absolute upper
limit on the rotation frequency, whereas other (less certain)
mechanisms, such as gravitational radiation reaction instabilities,
could impose lower limits on the maximal rotation frequency of a star.
Because the centrifugal potential counteracts the compressing stress
exerted on matter by gravity, rotating configurations are more massive
than their non-rotating counterparts. New-born hybrid stars can be
formed with large internal circulation, but viscosity and magnetic
stress act to dampen such motions at least in non-superfluid compact
stars.  Therefore, we can assume that cold single-component stellar
configurations are uniformly rotating to a good approximation.

These considerations motivate our study of the compact stars with
quark cores in a range of rotation periods from the Keplerian limit
down to the static limit. There are evolutionary considerations, which
make such a study physically relevant: Compact stars that are born
rapidly rotating evolve adiabatically by losing energy to
gravitational waves and electromagnetic (to lowest order, dipole)
radiation. This  evolution is well represented by star sequences with
constant rest mass and varying spin. The most massive members of
rapidly rotating configurations may not have a stable static limit;
that is, the evolution may terminate with a formation of a black
hole. 
These configurations are known as supramassive
configurations~\citep{1994ApJ...422..227C}.  The spin-up of a pulsar to
millisecond frequencies via accretion from a companion involves
changes in the rest mass and the spin of the star;  that is, simple
one-parameter evolutionary scenarios cannot be constructed without
additional physical input (\eg, mass accretion rate,
etc). Nevertheless, we are able to draw the boundary of
the region in this case,  where stable configurations with given spin and mass can
exist. The evolution of supramassive configurations were studied in
detail in \cite{1994ApJ...422..227C}  for a large set of
EoS for baryonic matter by using an exact numerical integration
method for Einstein's equations. 

The spin evolution of hybrid stars, or stars with quark-matter cores
surrounded by a nuclear envelope,  may differ from the evolution of
purely nucleonic stars.  Because the spin-down compresses the
matter within the star, it can induce a phase transition to the 
quark-matter phase once the critical density of the phase transition has
been reached~\citep{1997PhRvL..79.1603G,2000A&A...357..968C,2001ApJ...559L.119G,2002A&A...395..151S,2003NuPhA.715..831G,2006A&A...450..747Z,2009MNRAS.396.2269D,2009MNRAS.392...52A,2009A&A...502..605H,2013arXiv1307.1103W}. 

Intense studies of stationary (nonrotating) hybrid compact stars began
after the measurement of two solar mass pulsars PSR J1614-2230 
\citep{2010Natur.467.1081D}  to reconcile the phase transition from baryonic matter to quark
matter with these observations \
\citep{2011A&A...536A..92B,2011arXiv1112.6430L,2011PhRvD..84j5023C,2011RAA....11..687L,2012A&A...539A..16B,2012ApJ...759...57L,2012arXiv1212.1388C,2013arXiv1301.2675S,
2013ApJ...764...12M,
2013arXiv1307.0681S,
2013arXiv1307.1928L,
2013arXiv1307.6996K,
2013PTEP.2013g3D01M,
2013PhRvC..87a5804D,
2013arXiv1307.3995C}.
Massive stable compact stars with color-superconducting cores were obtained
only in a few studies prior to these
observations~\citep{2005ApJ...629..969A,2008PhRvD..77b3004I,2009PhRvD..79h3007K,2009PhRvD..79j3006A,2010PhRvD..81j5021K}.
Rotating configurations of color-superconducting stars were previously
constructed by \cite{2008PhRvD..77b3004I}; however, evolutionary
aspects of the problem were not discussed so far.

In this work, we explore the full parameter space that characterizes
rotating hybrid configurations on the basis of a carefully chosen EoS,
which permits a phase transition from nucleonic matter to
color-superconducting quark matter at some pre-selected density.  The
nucleonic EoS is based on a relativistic density functional of nuclear
matter with density-dependent (DD) couplings, according to
parametrization of \cite{2005PhRvC..71b4312L}. This parameterization
was recently used in \cite{2013PhRvC..87e5806C} to address the
hyperonization puzzle in compact stars, and our approach follows the
one adopted in that work.  The quark-matter EoS is based on the
phenomenological four-fermion interaction Nambu--Jona-Lasinio
Lagrangian, that is supplemented with repulsive vector interactions
and that is the same as that used by \cite{2012A&A...539A..16B}.  The
energy of quark matter is minimized by allowing the
color-superconducting phases of two- and three-flavor quark matter.
Our construction of the EoS has two key parameters: one is the onset
density of the phase transition from nucleonic to quark matter and the
second is the strength of the repulsive vector interactions.

Solutions of Einstein's equations for rotating, axially
symmetric configurations have been obtained by a number of authors
using either perturbative
methods~\citep{1967ApJ...150.1005H,1968ApJ...153..807H,1968Afz.....4..551S,1968Ap......4...87S,1971AZh....48..496A}
or direct numerical
integration~\citep{1989MNRAS.239..153K,1989MNRAS.237..355K,1993A&A...278..421B,1994ApJ...422..227C,1995ApJ...444..306S,1998A&AS..132..431N,1998PhRvD..58j4020B}.
For a review and further references, see \cite{lrr-1998-8}. The
computations reported below are based on a direct numerical integration
of the Einstein's equations and were carried out using the public
domain RNS code {(\tt www.gravity.phys.uwm.edu/rns/)}.

This paper is structured as follows. In Sec.~\ref{sec:EoS,} we discuss
the input EoS of hybrid stars. Section~\ref{sec:results} contains our
results for the sequences of non-rotating and rapidly rotating hybrid
stars. Rotational properties of evolutionary sequences are discussed
in Sec.~\ref{sec:Rotational}.  In Sec.~\ref{sec:PhaseTrans}, we discuss
the rotationally induced phase transition to the color-superconducting
state of quarks and its implications for the physics of hybrid stars.  Our
conclusions are collected in Sec.~\ref{sec:conclusion}.

\section{Equation of state}
\label{sec:EoS}

At the fundamental level, zero-temperature, $\beta$-equilibrated matter
is described in terms of two equations of state for nucleonic matter
(at low densities) and quark matter (at high densities). The EoSs of
these phases that are used in our study were introduced earlier
\citep{2012A&A...539A..16B}, whereas the nucleonic EoS has been
improved further by \cite{2013PhRvC..87e5806C}, where a density
functional with DD couplings has been implemented.  Here, we combine
these EoS to describe hybrid-star matter by assuming a first-order phase
transition from nucleonic to quark matter.  Our approach allows quark
matter to support two types of color-superconducting phases: the
low-density phase is the two-flavor color-superconducting phases
(hereafter 2SC), and the high-density phase is the three-flavor
color-flavor-locked (hereafter CFL) phase. 
Readers interested in detailed discussions of color superconductivity 
should consult the reviews 
\citep{1984PhR...107..325B,
2000hep.ph...11333R,
2001ARNPS..51..131A,2004PrPNP..52..197R,
2005FoPh...35.1309S,2008RvMP...80.1455A,2013arXiv1302.4264A}.

Because the physics
underlying these EoSs has been described elsewhere, we provide only a
brief overview of these EoSs to keep our discussion self-contained.

\subsection{Nuclear matter}
\label{subsec:A}

Nucleonic matter is described by the relativistic Lagrangian,
\bea\label{eq:lagrangian} \nonumber {\cal L}_H & = &
\sum_N\bar\psi_N\bigg[\gamma^\mu \left(i\partial_\mu-g_{\omega
    N}\omega_\mu
  - \frac{1}{2} g_{\rho N}\tauvec\cdot\rhovec_\mu\right)\\
\nonumber & - & (m_N - g_{\sigma N}\sigma)\bigg]\psi_N
+ \frac{1}{2} \partial^\mu\sigma\partial_\mu\sigma-\frac{1}{2} m_\sigma^2\sigma^2\\
\nonumber & - & \frac{1}{4}\omega^{\mu\nu}\omega_{\mu\nu} +
\frac{1}{2} m_\omega^2\omega^\mu\omega_\mu -
\frac{1}{4}\rhovec^{\mu\nu}\rhovec_{\mu\nu}
+ \frac{1}{2} m_\rho^2\rhovec^\mu\cdot\rhovec_\mu\\
& + & \sum_{\lambda}\bar\psi_\lambda(i\gamma^\mu\partial_\mu -
m_\lambda)\psi_\lambda - \frac{1}{4}F^{\mu\nu}F_{\mu\nu}, \eea where
the $N$-sum is over the nucleons, $\psi_N$ are the nucleonic Dirac
fields with masses $m_N$, the $\lambda$-sum runs over the leptons
$e^-,\mu^-,\nu_e$, and $\nu_\mu$ with masses $m_\lambda$ and the last
term is the electromagnetic energy density.  The interaction among the
nucleons is mediated by the $\sigma,\omega_\mu$, and $\rhovec_\mu$
meson fields with $\omega_{\mu\nu}$ and $\rhovec_{\mu\nu}$ being the
corresponding field strength tensors and with masses $m_{\sigma}$,
$m_{\omega}$, and $m_{\rho}$.  The nucleon ($N$) and meson coupling
constants ($g_{iN}$) are density-dependent:
\bea
g_{iN}(\rho_N) &=& g_{iN}(\rho_{0})h_i(x), \qquad i=\sigma,\omega, \\
g_{\rho N}(\rho_N) &=& g_{\rho N}(\rho_{0})\exp[-a_\rho(x-1)], 
\eea
where $\rho_N$ is the nucleon density, $x=\rho_N/\rho_{0}$, $\rho_{0}$
is the nuclear saturation density, and 
\be\label{ansatz} h_i(x) =
a_i\frac{1+b_i(x+d_i)^2}{1+c_i(x+d_i)^2} .  
\ee 
We use the DD-ME2 parameterization of \cite{2005PhRvC..71b4312L}. The
parameter values listed in Table~\ref{table:DD-ME2} are adjusted to
reproduce the properties of symmetric and asymmetric nuclear matter,
binding energies, charge radii, and neutron radii of spherical nuclei.

\begin{table}[ht]
 \centering
\caption{Meson masses and couplings to the nucleons in the DD-ME2 effective interaction.}
\label{table:DD-ME2}
 \begin{tabular}{cccc}
\hline
          &     $ \sigma$    & $\omega$ & $\rho$  \\
\hline
   $m_i$ [MeV] &  550.1238  &  783.0000    &  763.0000  \\
   $g_{Ni}(\rho_{0})$ & 10.5396    & 13.0189     &  3.6836\\
   $a_i$  & 1.3881    & 1.3892 &   0.5647\\
   $b_i$  & 1.0943   & 0.9240 &   ---\\
   $c_i$  & 1.7057   &  1.4620 &   ---\\
   $d_i$  & 0.4421  & 0.4775 &    ---\\
\hline
 \end{tabular}
\end{table}

The Lagrangian density (\ref{eq:lagrangian}) yields the
zero-temperature pressure of nucleonic matter   as a
function of density 
\bea\label{eq:thermodynamic}
P_H& = & - \frac{m_\sigma^2}{2} \sigma^2 + \frac{m_\omega^2}{2}
\omega_0^2 + \frac{m_\rho^2 }{2} \rho_{03}^2
+ \frac{1}{3}\sum_N \frac{2J_N+1}{2\pi^2} \nonumber\\
&\times &\int_0^{\infty}\!\!\! \frac{k^4 \ dk}{(k^2+m^{* 2}_N)^{1/2}}
\left[\theta(-E_k^B+\mu_N^*)+\theta(-E_k^B-\mu_N^*)\right]\nonumber\\
& +& \frac{1}{3\pi^2} \sum_{\lambda} \int_0^{\infty}\!\!\!  \frac{k^4
  \
  dk}{(k^2+m^2_\lambda)^{1/2}}\nonumber\\
&\times&
\left[\theta(-E_k^{\lambda}+\mu_\lambda)+\theta(-E_k^{\lambda}-\mu_\lambda)\right] +
\rho_N\Sigma_{r}, \eea where $\sigma$, $\omega_0$, and $\rho_{03}$ are
the nonvanishing mesonic mean fields, $J_N$ is the nucleon spin,
 $m_N^* = m_N - g_{\sigma B}$ is the effective nucleon mass,
$\mu^*_N = \mu_N-g_{\omega N}\omega_0 - g_{\rho N}I_3\rho^0_3$ is the effective
baryon chemical potential, $I_3$ is the third component of baryon isospin, $E_k^B =
\sqrt{k^2+m^{*2}_N}$ and $E_k^{\lambda}=\sqrt{k^2+m_\lambda^2}$ are
the single-particle energies of baryons and leptons (electrons $e$ and
muons $\mu$), respectively; and $\theta(x)$ is the step function. The
lepton masses $m_\lambda$ $(\lambda \in e, \mu)$ are taken to be equal
to their free-space values.

\subsection{Quark matter}
\label{subsec:B}

Quark matter is described by an extended Nambu--Jona-Lasinio Lagrangian:
\begin{eqnarray}
\label{eq:NJL_Lagrangian}
\mathcal{L}_{Q}&=&\bar\psi(i\gamma^{\mu}\partial_{\mu}-\hat m)\psi 
+G_S \sum_{a=0}^8 [(\bar\psi\lambda_a\psi)^2+(\bar\psi i\gamma_5 \lambda_a\psi)^2]\nonumber\\
&+& G_D \sum_{\gamma,c}[\bar\psi_{\alpha}^a i \gamma_5
\epsilon^{\alpha\beta\gamma}\epsilon_{abc}(\psi_C)^b_{\beta}][(\bar\psi_C)^r_{\rho} 
i \gamma_5\epsilon^{\rho\sigma\gamma}\epsilon_{rsc}\psi^8_{\sigma}]\nonumber\\
&-&K \left \{ {\rm det}_{f}[\bar\psi(1+\gamma_5)\psi]+{\rm
    det}_{f}[\bar\psi(1-\gamma_5)\psi]\right\}\nonumber\\
&+&G_V(\bar\psi i \gamma^{\mu}\psi)^2,
\end{eqnarray}
where the quark spinor fields $\psi_{\alpha}^a$ carry color ($a = r, g,
b$) and flavor ($\alpha= u, d, s$) indices, the matrix of the quark current
masses is given by $\hat m= {\rm diag}_f(m_u, m_d, m_s)$, $\lambda_a$
with $ a = 1,..., 8$ are the Gell-Mann matrices in color space,
and $\lambda_0=(2/3) {\bf 1_f}$.  The charge-conjugate spinors are
defined as $\psi_C=C\bar\psi^T$ and $\bar\psi_C=\psi^T C$, where
$C=i\gamma^2\gamma^0$ is the charge conjugation matrix.  The standard
Nambu--Jona-Lasinio Lagrangian is extended here to include vector interactions
($\propto G_V$) among quarks and the 't Hooft interaction term
($\propto K$).

The pressure of quark matter at zero temperature follows from the
saddle-point evaluation of the partition function of quark matter as
described by the Lagrangian~(\ref{eq:NJL_Lagrangian}),
\begin{eqnarray}
P_Q&=&\frac{1}{2\pi^2}\sum_{i=1}^{18}\int_{0}^{\infty}dk k^2
\vert\epsilon_i\vert \theta(\Lambda-\vert k_{Fi}\vert) \theta(k_{Fi}-\vert k\vert)
\nonumber\\
&+&4 K \sigma_u\sigma_d\sigma_s
-\frac{1}{4G_D}\sum_{c=1}^{3}\vert\Delta_c\vert^2
-2G_s\sum_{\alpha=1}^{3}\sigma_{\alpha}^2\nonumber\\
&+&\frac{1}{4
  G_V}(2\omega_0^2+\phi_0^2)-P_0-B^*,
\end{eqnarray}
where $\epsilon_i$ are the quasiparticle spectra of quarks, $k_{Fi}$
their Fermi-momenta, $\Lambda$ is the cut-off, $\omega_0=G_V\langle QM
\vert \psi_u^{\dagger}\psi_u+\psi_d^{\dagger}\psi_d\vert QM\rangle$
and $\phi_0=2 G_V\langle QM \vert \psi_s^{\dagger}\psi_s\vert
QM\rangle$ are the mean-field expectation values of the vector mesons
$\omega$ and $\phi$ in quark matter, $P_0$ is the vacuum pressure, and
$B^*$ is an effective bag constant. The first $\theta$-function
guarantees that the Fermi sphere of quarks lies within the momentum
space spanned by the model, which is limited by the cut-off $\Lambda$.
In the 2SC phase, leptons contribute to the pressure of quark matter
and, as in the case of baryonic matter, we assume that the leptons
form an ideal gas.  The quark chemical potentials are modified by the
vector fields as follows: $\mu_Q^*={\rm
  diag}_f(\mu_u-\omega_0,\mu_d-\omega_0,\mu_s-\phi_0)$.  The numerical
values of the parameters of the Lagrangian are $m_{u,d} = 5.5$ MeV,
$m_s = 140.7$ MeV, $\Lambda = 602.3$ MeV, $G_S\Lambda^2 = 1.835$,
$K\Lambda^5 =12.36$, and $G_D/G_S = 1$. The strength of the vector
coupling $G_V$ and the transition density from nucleonic matter to
quark matter are free parameters of our model.

\subsection{Matching the equations of state}

Because the surface tension between nuclear and quark matter is not
well determined, the matching between the nuclear and quark EoS can be
carried out in two possible ways.  If the tension between these phases
is low, then a mixed phase of quark and nuclear matter forms. The
second possibility is that the transition boundary is sharp, which is
the case when there is  sufficiently high tension between the phases. In this
case, the transition occurs at a certain baryo-chemical potential at
which the pressures of the two phases become equal. This is equivalent
to the condition that curves of pressure, $P$, vs. baryo-chemical
potential, $\mu$ for these phases cross; that is, the phase equilibrium is
determined by a Maxwell construction.  This implies (according to the
standard Maxwell construction) that at the deconfinement phase
transition, there is a jump in density at constant pressure.  Because
the transition is first order, latent heat is released during the
transition.  However, the transition density itself cannot be fixed,
because the Nambu--Jona-Lasinio model does not allow us to fix the
low-density normalization of the pressure (this is a consequence of
the fact that this class of models does not capture the confinement
feature of QCD). Therefore, we introduced an additional
``bag'' parameter $B^*$ above, which allows us to vary the density at which
the quark phase sets in, thus fixing the density of deconfinement
$\rho_{\rm tr}$.

\begin{figure}[t]
\begin{center}
 \includegraphics[height=7cm,width=7cm,angle=0]{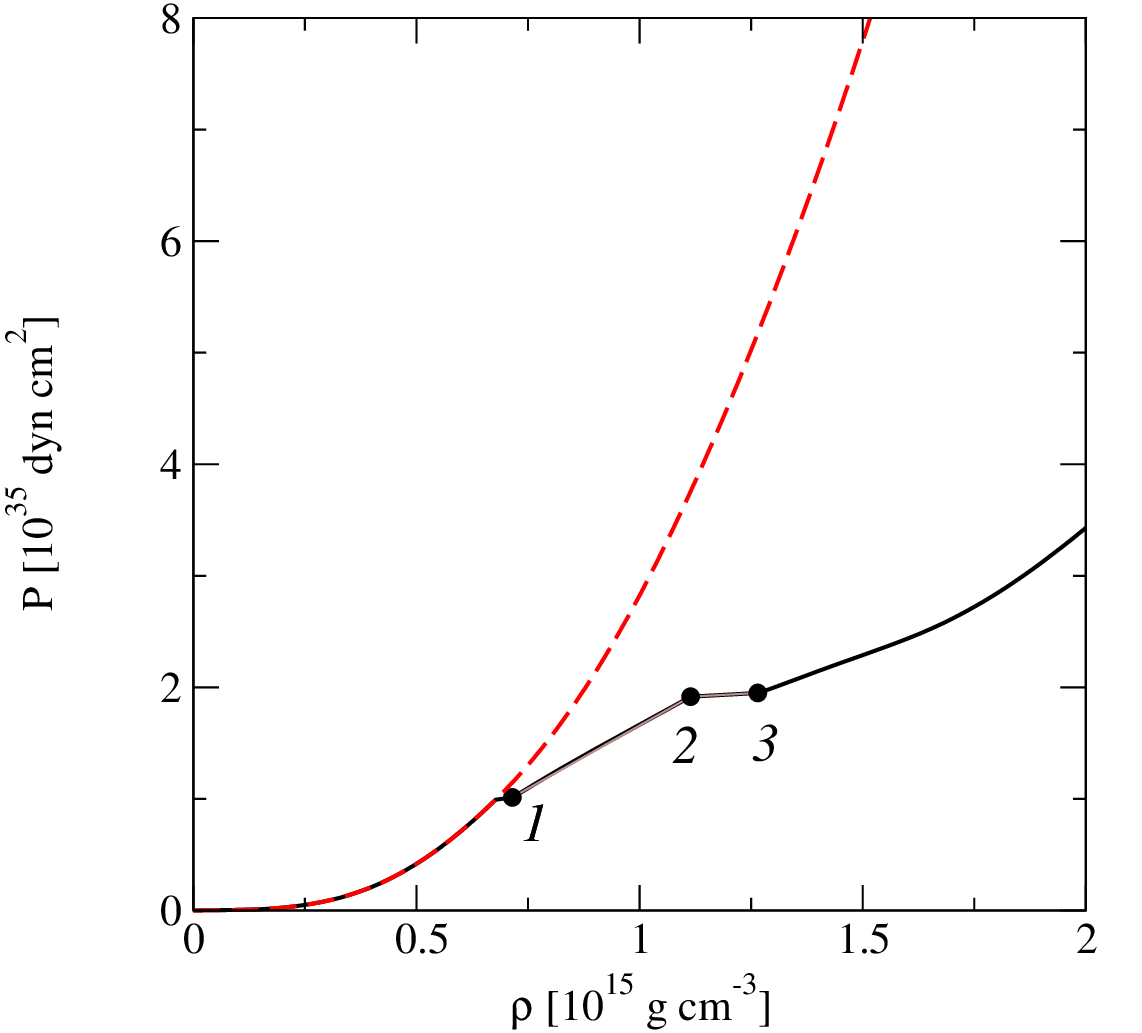}
\end{center}
\caption[]
{ The EoS of nucleonic matter (dashed, red line) and of
 hybrid matter (solid, black line). The interval between the points 1 and 2
 corresponds to the 2SC phase, whereas the interval to the right
 from  point 3 corresponds to the CFL phase. 
}\label{fig:EOS}
\end{figure}

The resulting EoS of pure nucleonic and nucleonic plus quark matter
is shown in Fig.~\ref{fig:EOS}. For brevity we shall refer to these
EoS as EoS A and EoS B, respectively. The EoS B displays two (first-order)
transitions, one from nuclear matter to two-flavor
color-superconducting matter (2SC phase) and the second from the 2SC
phase to the color-flavor-locked (CFL) phase. The segment of the EoS B
between points 1 and 2 in Fig.~\ref{fig:EOS} corresponds to the 2SC
phase; the segment to the right from point 3 corresponds to the CFL
phase.  In the following, we fix the values of the two parameters
$\rho_{\rm tr}=2.5\rho_0$, where $\rho_0$ is the nuclear saturation
density and the ratio $G_V/G_S = 0.8$.

\section{Mass, radius, and moment of inertia}
\label{sec:results}

\label{sec:NR-confug}
\begin{figure}[tb]
\begin{center}
 \includegraphics[height=7.0cm,width=7cm,angle=0]{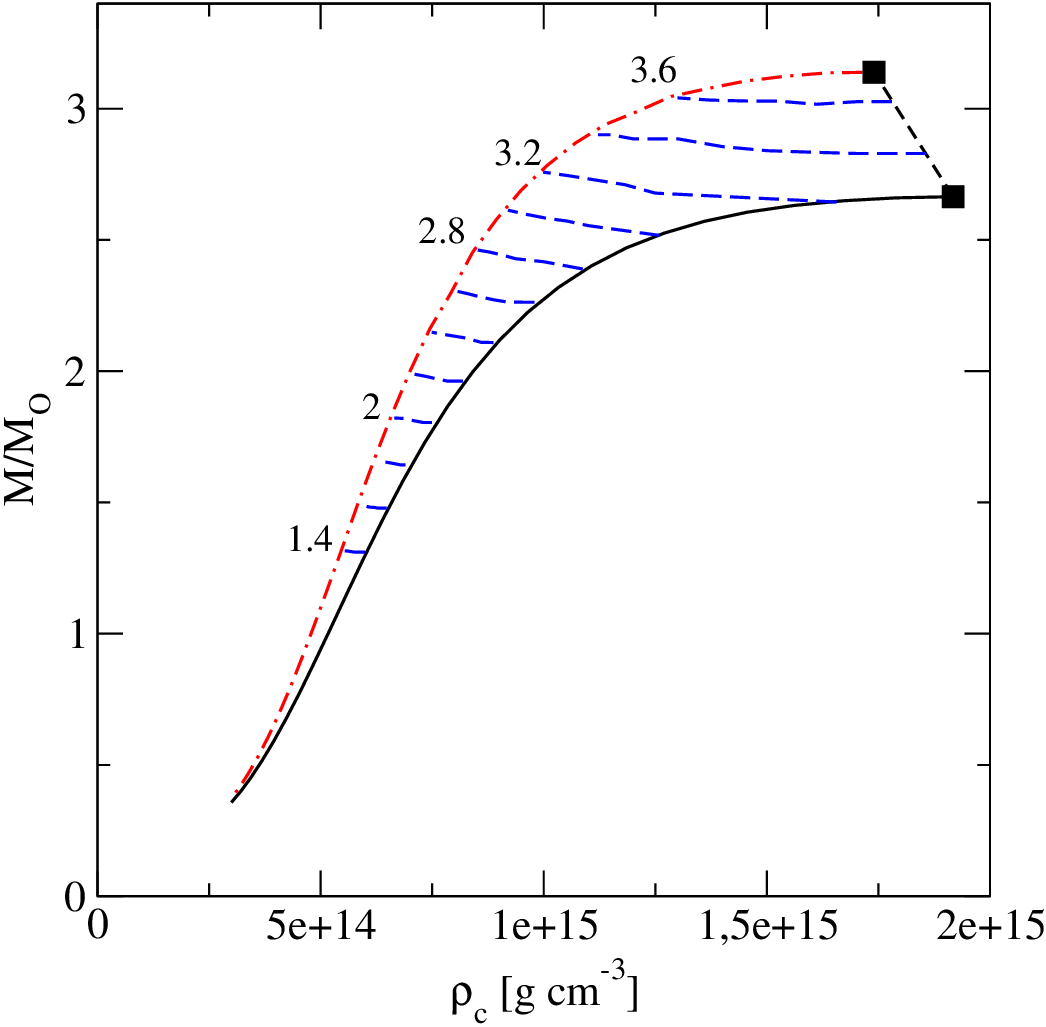}
 \vskip 1.2cm
 \includegraphics[height=7.0cm,width=7cm,angle=0]{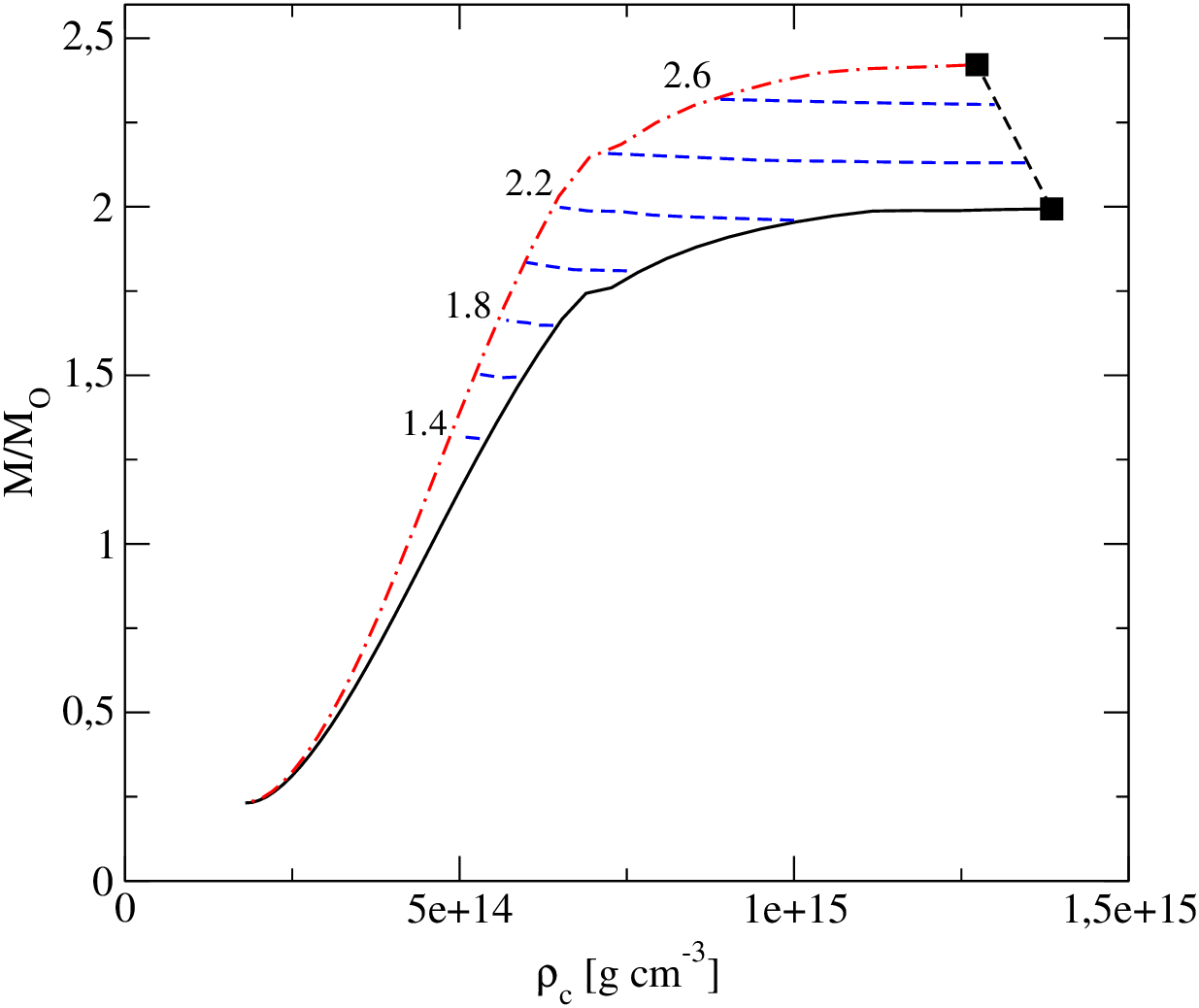}
%\includegraphics[keepaspectratio,width=0.957cm,angle=0]{nucM_vs_Ec.eps}
%\vskip 1cm
%\includegraphics[keepaspectratio,width=7cm,angle=0]{M_vs_Ec.eps}
\end{center}
\caption[] { Dependence of masses of hybrid compact
 stars in solar units on their central density for EoS A (upper
 panel) and EoS B (lower panel). Solid (black online) lines show the
 non-rotating sequence. The dash-dotted (red online) lines show
 the maximally fast rotating sequence. The horizontal dashed (blue online) lines
 show constant rest-mass sequences, some of which are labeled by the
 value of their rest mass (in solar units).  The mass step between
 two sequential evolutionary sequences is 0.2 $M_{\odot}$.
 }
\label{fig:mass_cdens}
\end{figure}

A well-known feature of the static (spherically-symmetric) or
uniformly rotating (axially symmetric) equilibrium configurations in
General Relativity is the existence of a maximum mass for sequences
parameterized by their central density $\rho_c$. This feature is
independent of any particular EoS and is illustrated for the sequences
based on the EoS A and B in Fig.~\ref{fig:mass_cdens}.  If the central
density of a configuration is larger than the value corresponding to
the maximum mass, it becomes unstable toward collapse to a black
hole. This condition is a sufficient condition for a secular
instability in uniformly rotating stars~\citep{1988ApJ...325..722F}.
A necessary criterion for instability is the neutrality (zero) of the
frequency of the fundamental (pulsation) modes. The turning points,
where $dM/d\rho_c = 0$, and the neutrality lines of fundamental
frequency modes are close to each other for models of rotating
configurations, but they do not coincide~\citep{2011MNRAS.416L...1T}.
Thus, some uniformly rotating configurations that are stable according
to the turning point criterion can be still unstable either
dynamically or secularly.  However, one can still use the turning
point criterion for practical purpose of locating the instability of
uniformly rotating stars.

Our purely nuclear EoS belongs to the class of hard equations of
states based on relativistic density functionals; therefore, we find a
large maximum mass for the purely nucleonic sequences, which is on the
order of 2.7~$M_{\odot}$ (Fig.\ref{fig:mass_cdens}, upper
panel). Hybrid configurations branch off from the nuclear
configurations when the central density of a configuration reaches the
density of the deconfinement phase transition
(Fig.\ref{fig:mass_cdens}, lower panel). The jump in the density at
constant pressure is translated into a plateau in the $M(\rho_c)$
dependence, where $dM/d\rho_c = 0$. At the maximal value of $\rho_c$,
where this plateau ends, the hybrid stars emerge.  

Equilibrium configurations with central densities
  located within the plateau cannot exist, because 
 the gravitational force cannot be balanced if there are no
  pressure gradients. Indeed, $dP/dr = (dP/d\rho) (d\rho/dr) = 0$,
  $d\rho/dr\neq 0$, and $dP/d\rho = 0$ at any point on the plateau
  (see Fig.~\ref{fig:EOS}). Thus, there are two stable families of
  purely hadronic stars and hybrid stars on the left and right sides
  of the plateau, respectively, but there are no equilibria
  in-between. Any perturbation of the central density of a
  configuration, which  results in a new central density located on
  the plateau makes the central region of the star unstable. This does
  not necessarily imply that the star would become entirely unstable.

 Dynamical processes, such as accretion, collapse, or
  secular spin-down, may result in a transient configuration with
  a central density located on the plateau.  In this case, a
  rearrangement of matter should take place such that the pressure
  gradients are restored, and the central density of the configuration
  is outside of the plateau. Thus, for example, catastrophic
  rearrangement of matter in accreting systems may take place because
  of gradual compression. This occurs once the transition density
  from hadronic to the color-superconducting quark matter is reached.
  These rearrangements are accompanied with an energy release equal to
  the difference in the energies of configurations located at the
  end points of the plateau.  Since the plateau also appears in the
  case of rotating configurations, the arguments above are valid for
  rotating stars as well. Furthermore, similar arguments apply to the
  transition from the purely 2SC core to the 2SC plus CFL core hybrid
  stars, which are separated by another plateau  (see Fig.~\ref{fig:EOS}).

The non-rotating sequences terminate at the maximum mass, where
stability is lost. Rotating nucleonic and hybrid compact stars cover
an area in Fig.~\ref{fig:mass_cdens}, which is bounded from above by
the mass-shedding configurations (dash-dotted lines) and from below by
the static configurations. These static configurations are simply the
solutions of the  spherically symmetric
Tolman-Oppenheimer-Volkoff equations. In our plots, these are
connected to their maximally fast-rotating counterparts via their
evolutionary sequences, which corresponds to the rotational evolution
at constant {\it rest-mass} (almost horizontal dashed curves). The
rest masses of evolutionary sequences were computed with a mass step
$0.2 M_{\odot}$.  As classified by \cite{1994ApJ...422..227C}, the
constant rest-mass sequences naturally fall into two classes: the
normal sequences, which always have a stable static (non-rotating)
limiting configuration, and the supramassive sequences, which do not
have such limit. These sequences terminate on the line that joins the
maximal masses in the static and Keplerian limits, which thus provides
a boundary of the area of stable configurations in the upper
right-hand corner of Fig.~\ref{fig:mass_cdens}.  The area where stable
stars are possible is bounded from below by the minimum possible mass
of configurations; the latter bound is not of interest to us, because
the central densities of the minimal-mass configurations are well
below the phase-transition density.
\begin{figure}
\begin{center}
 \includegraphics[height=7cm,width=7cm,angle=0]{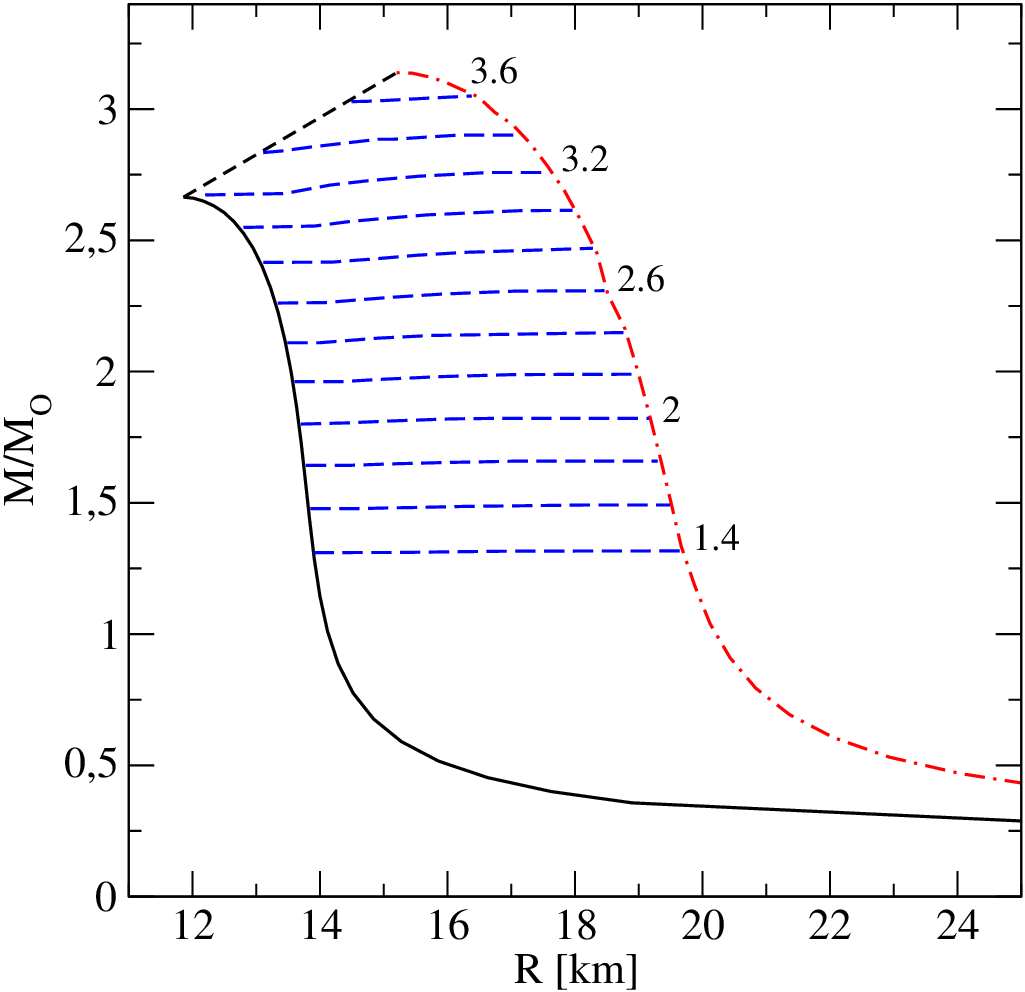}
 \vspace{1.cm}
 \includegraphics[height=7cm,width=7cm,angle=0]{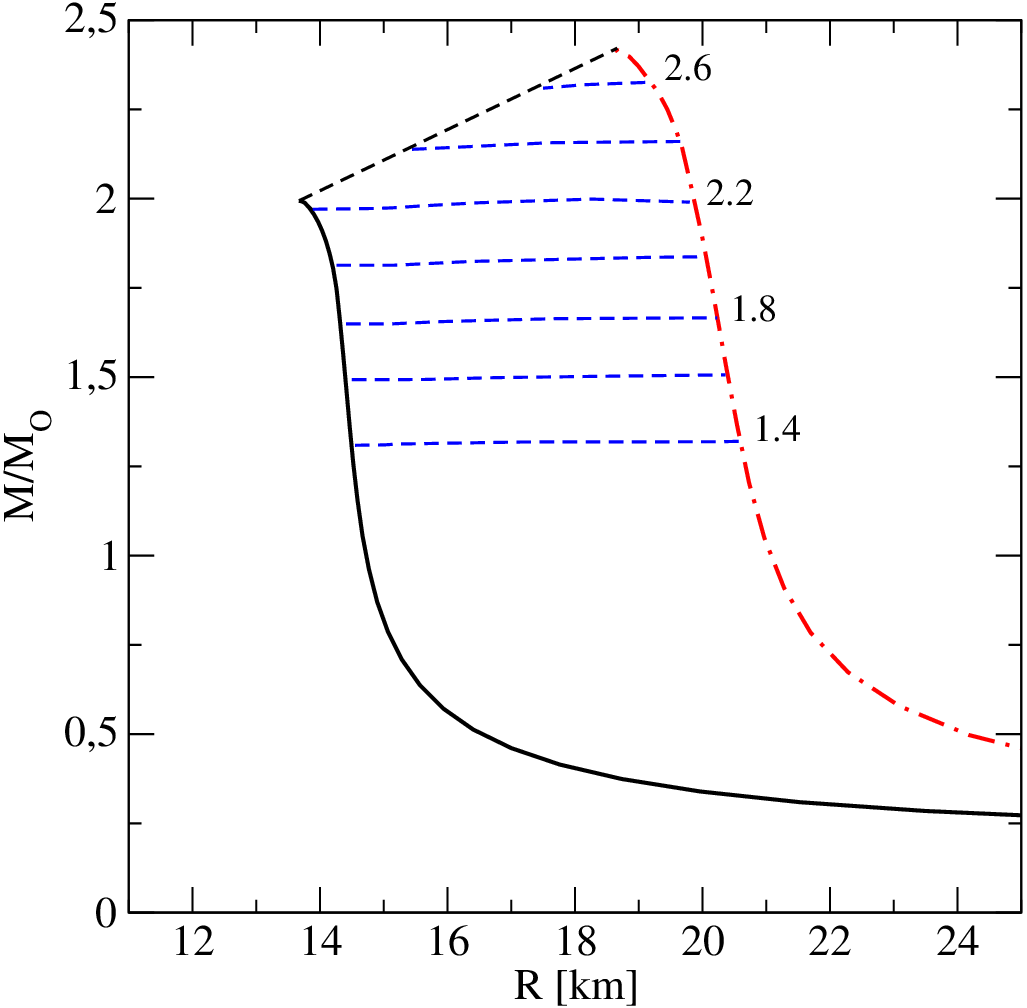}
%\includegraphics[keepaspectratio,width=7cm,angle=0]{nucM_vs_R.eps}
%\vspace{1.cm}
%\includegraphics[keepaspectratio,width=7cm,angle=0]{M_vs_R.eps}
\end{center}
\caption[]
{ The mass-radius relationship for hybrid compact
  stars  as described by  EoS A (upper  panel) and EoS B (lower
  panel). The labeling is the same as in Fig.~\ref{fig:mass_cdens}.
}\label{fig:MR_hybrid}
\end{figure}
Because the phase-transition density is independent of rotation
frequency and the rotating sequences lie above the static sequences,
there always exist evolutionary sequences, whose Keplerian limit
corresponds to purely nucleonic configurations, whereas the static
limit corresponds to a hybrid one (see Fig.~\ref{fig:mass_cdens}).
Consequently, the evolutionary sequences must involve a rotational
frequency range, which is characterized by {\it a phase transition to a
  color-superconducting quark state} in the interior of the star.  We
shall call such evolutionary sequences as {\it transitional
  sequences.}  It is clear from this figure that all supramassive
sequences are either transitional or contain quark matter already at
the Keplerian limit. However, not all transitional sequences are
supramassive; one such example is the $2.2 M_{\odot}$ rest-mass star
shown in Fig.~\ref{fig:mass_cdens}.  Therefore, it is not excluded
that slowly rotating hybrid stars that are close to the static limit
have undergone a phase transition to the superconducting state in the
course of the spin-down from the initial spin acquired at the birth.

Figure~\ref{fig:MR_hybrid} shows the gravitational masses of the
sequences as a function of equatorial circumferential radius for
static, maximally rotating, and evolutionary sequences. Along the
evolutionary sequences, the radius of a star shrinks to its limiting
value corresponding to the static configuration. For the transitional
sequences, this is accompanied by the onset of quark superconductivity
and growth of the radius of the quark core.
\begin{figure}
\begin{center}
 \includegraphics[height=7cm,width=7cm,angle=0]{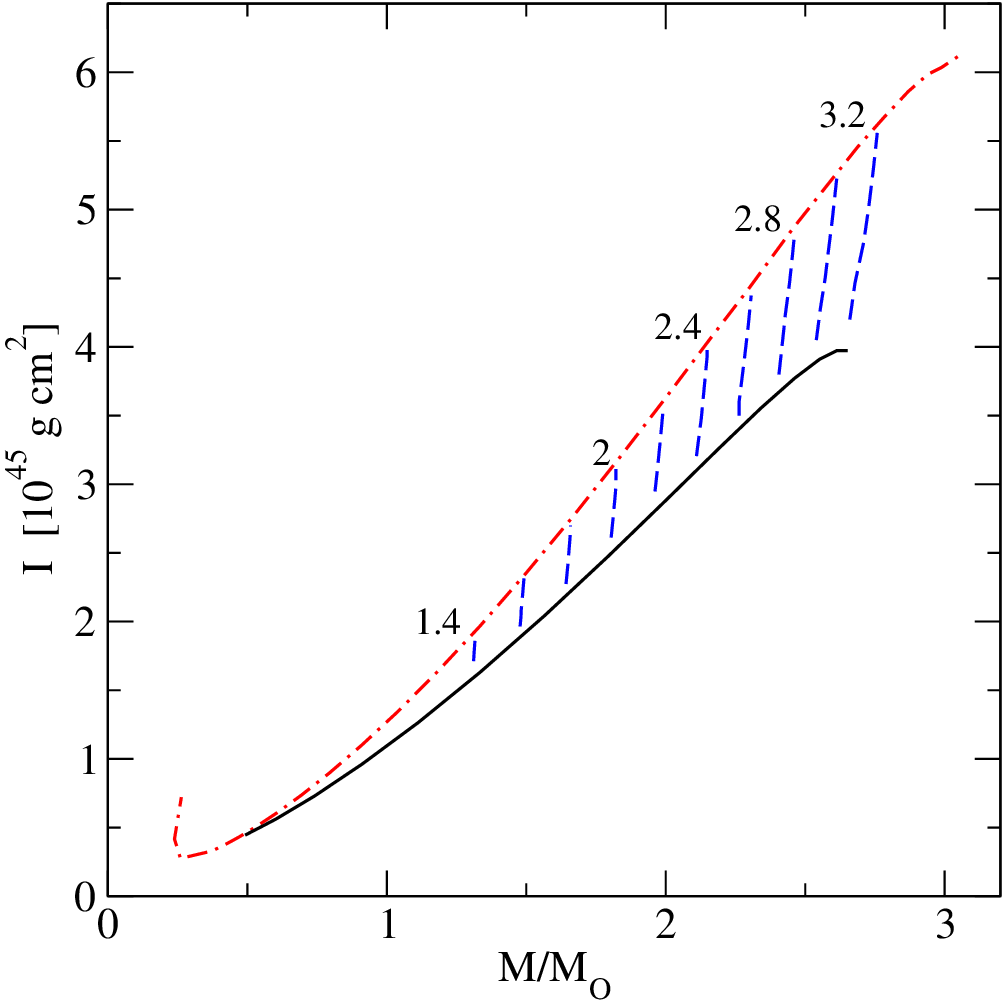}
 \vspace{1.cm}
 \includegraphics[height=7cm,width=7cm,angle=0]{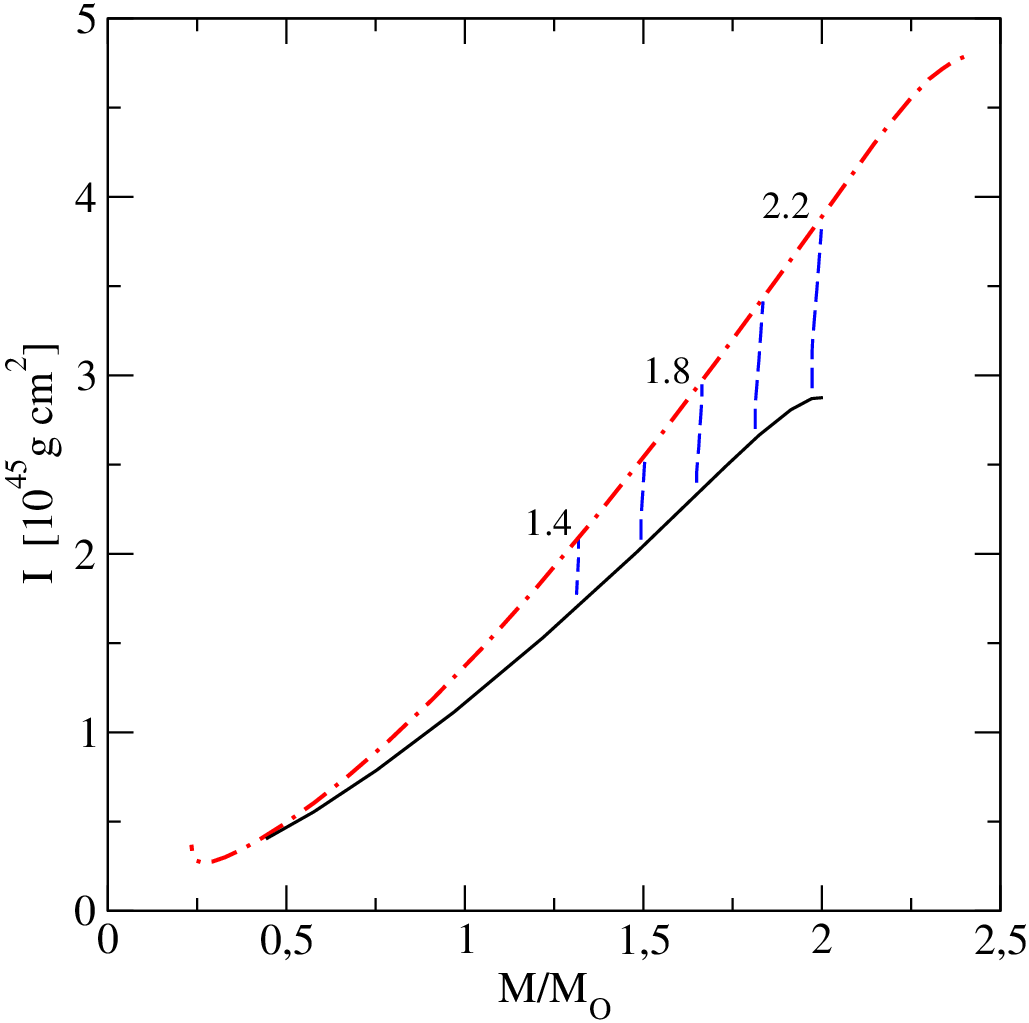}
%\includegraphics[keepaspectratio,width=0.947cm,angle=0]{nucI_vs_M.eps}
%\vspace{1.cm}
%\includegraphics[keepaspectratio,width=0.947cm,angle=0]{I_vs_M.eps}
\end{center}
\caption[]
{Dependence of moment of inertia of the hybrid configurations
on the mass of configuration (in solar mass units) for EoS A (upper
panel) and EoS B (lower panel). The labeling is the same 
as in Fig.~\ref{fig:mass_cdens}.
}\label{fig:mom_in}
\end{figure}
Figure~\ref{fig:mom_in} shows the dependence of moments of inertia of
the sequences on the masses of the stars. The decrease in the moments
of inertia from the value at the Keplerian to the value in the static
limit as one moves along the evolutionary sequences is the consequence
of the decreasing radius of a star at constant rest mass. This feature
is not affected by the emergence of superconducting quark matter in
the transitional sequences for uniformly rotating stars.

\section{Rotational evolution}
\label{sec:Rotational}

\begin{figure}
\begin{center}
% \includegraphics[height=7cm,width=7cm,angle=0]{nucOmega_vs_J.eps}
% \vspace{1.cm}
% \includegraphics[height=7cm,width=7cm,angle=0]{Omega_vs_J.eps}
\includegraphics[keepaspectratio,width=7cm,angle=0]{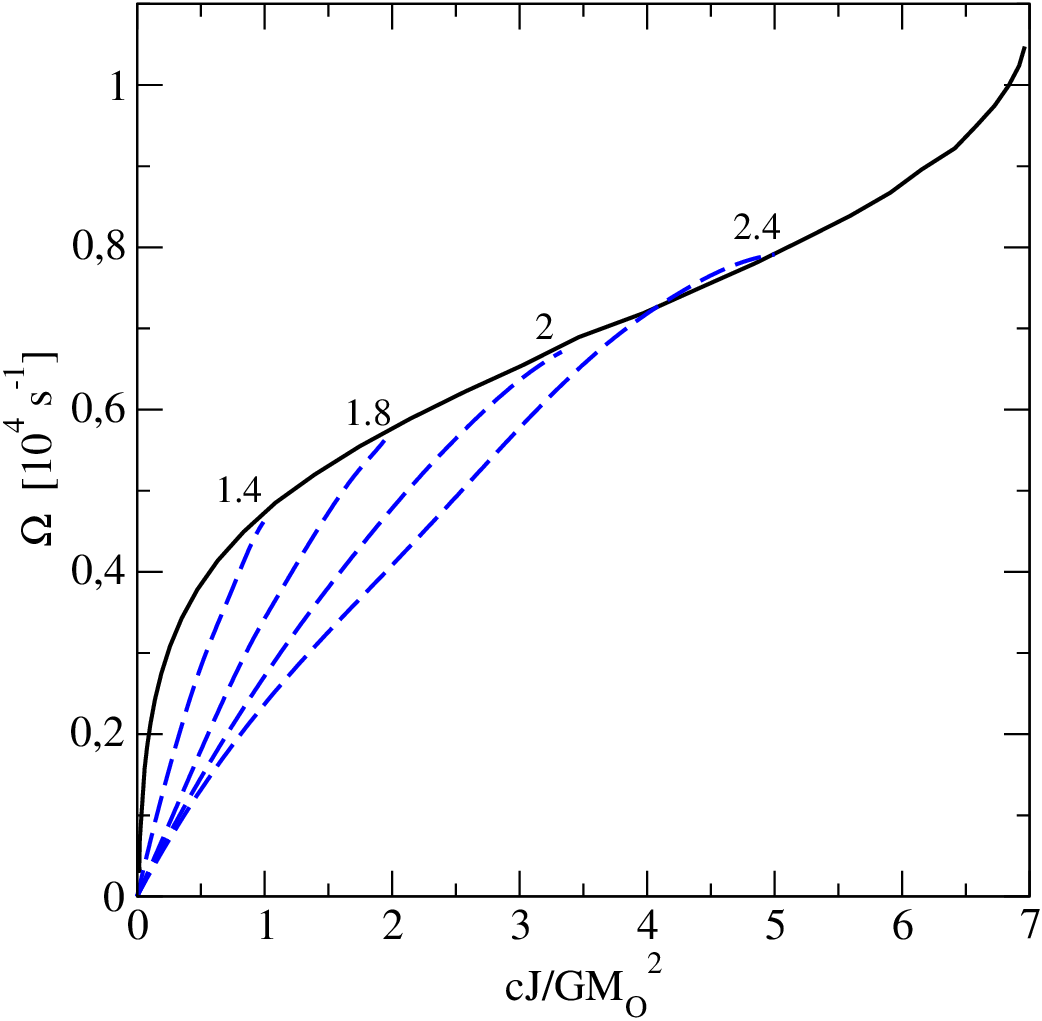}
\vspace{1.cm}
\includegraphics[keepaspectratio,width=7cm,angle=0]{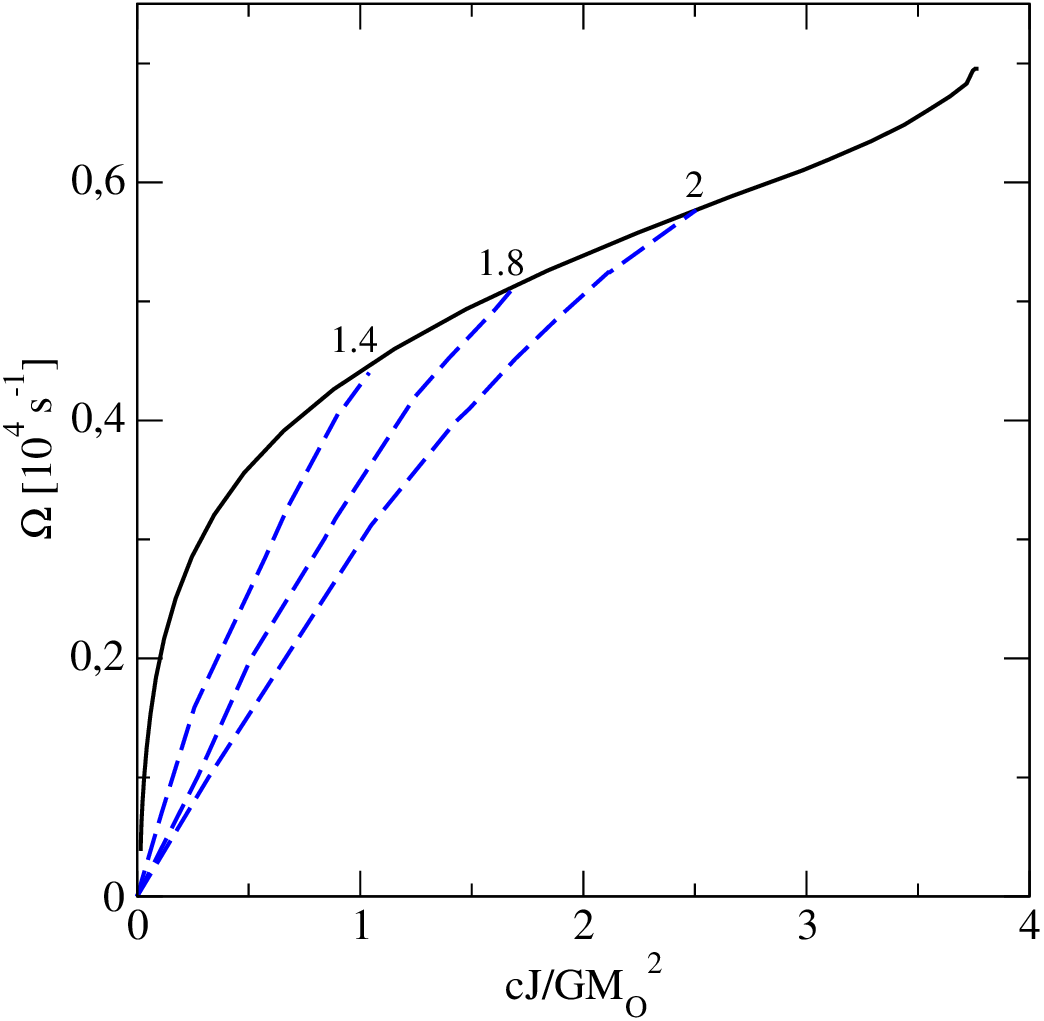}
\end{center}
\caption[]
{Dependence of spin frequency of configurations on the angular
  momentum. The constant rest mass evolutionary sequences are shown by
  dashed lines and the Keplerian sequence by the solid line. The 
low-$J$ results  ($cJ/GM_{\odot} \le 0.4)$ were obtained by
  interpolation. The supramassive sequences (not shown) spin-up as
  they approach the onset of instability. The upper panel corresponds
  to EoS A and the lower panel  to EoS B.
}\label{fig:Om_J}
\end{figure}

\begin{figure}
\begin{center}
% \includegraphics[height=7cm,width=7cm,angle=0]{nucOmega_vs_M.eps}
% \vspace{1.cm}
% \includegraphics[height=7cm,width=7cm,angle=0]{Omega_vs_M.eps}
\includegraphics[keepaspectratio,width=7cm,angle=0]{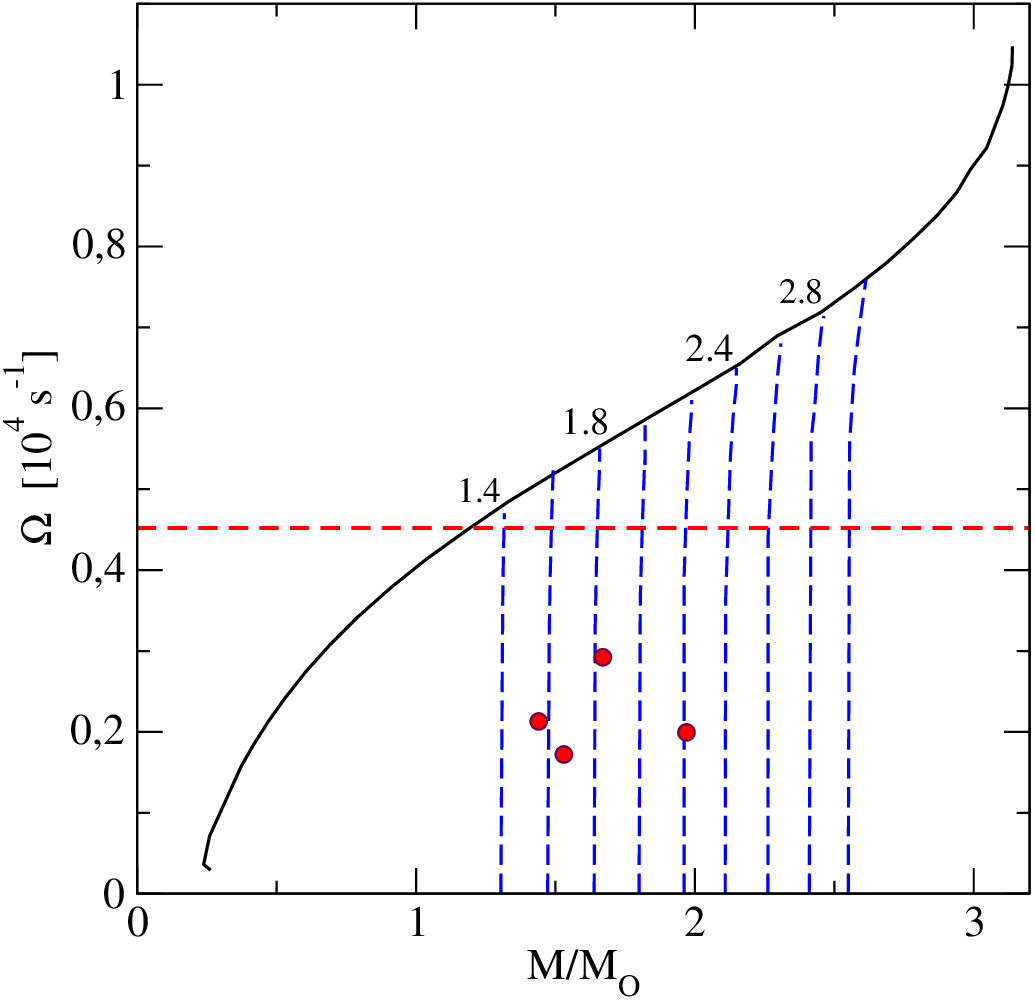}
\vspace{1.cm}
\includegraphics[keepaspectratio,width=7cm,angle=0]{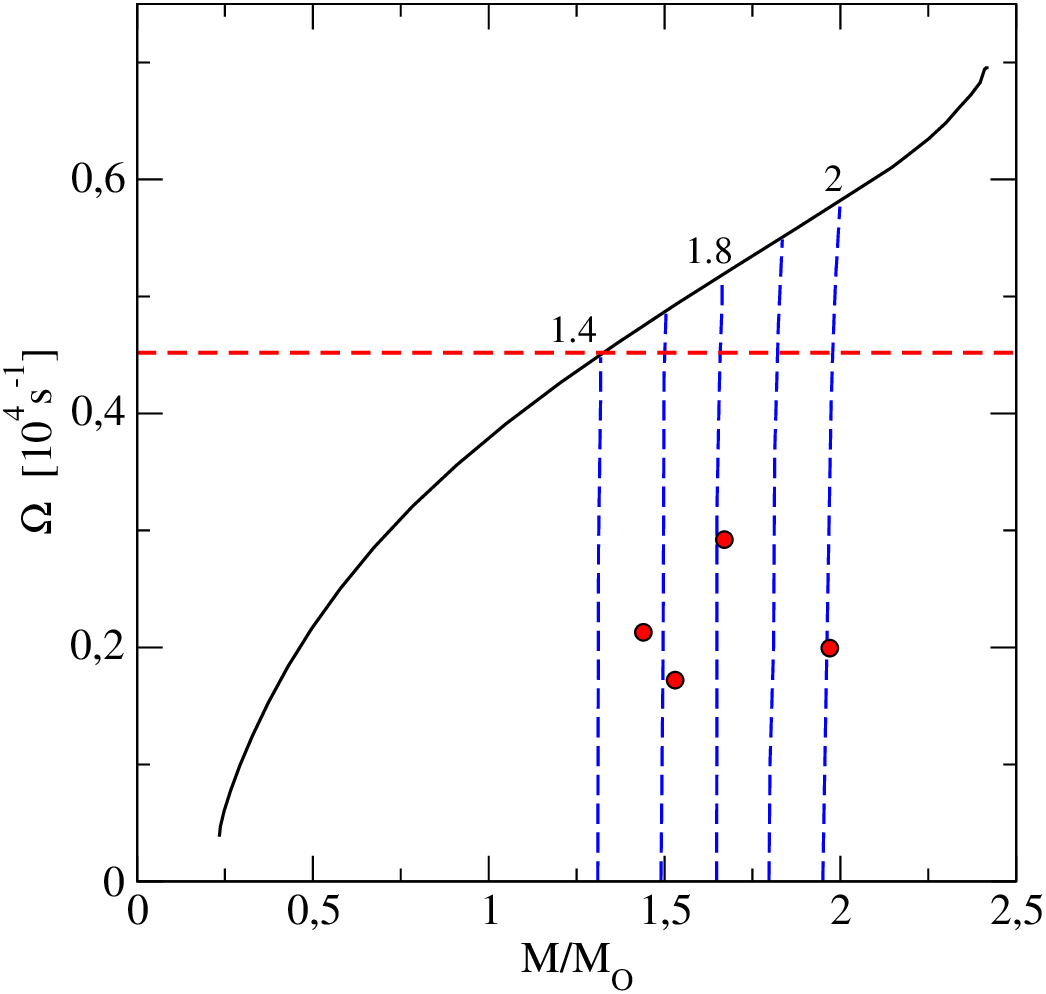}\end{center}
\caption[]
{
Dependence of spin frequency of configurations on  gravitational mass.
The conventions are the same as in Fig~\ref{fig:Om_J}. The dots show 
the four pulsars listed in Table~\ref{tab:PSR}, which are among fast rotating
pulsars with available mass measurements. The horizontal dashed
line is the spin frequency of the fastest rotating pulsar
J$1748-2446$.
The upper panel corresponds
  to EoS A and  the lower panel  to EoS B.
}\label{fig:Om_M}
\end{figure}

The variation of spin frequency along the normal evolutionary and
transitional sequences as a function of angular momentum of the star
is shown in Fig.~\ref{fig:Om_J}. For all masses, the stars belonging to
the normal evolutionary sequences spin down as they lose angular
momentum. The transitional sequence with the rest mass $2M_{\odot}$
shows the same feature, assuming that the quark superconductor rotates
uniformly with the hadronic matter. We verified that supramassive
sequences spin up as they lose angular momentum, which agree with the
results of \cite{1994ApJ...422..227C} as obtained for a number of
nucleonic EoSs. This effect is independent of the onset (or presence
from the outset) of quark matter in the cores of compact stars, as
long as they are assumed to rotate uniformly with the nucleonic
envelope.
Figure~\ref{fig:Om_M} displays the dependence of the rotation
frequency of a configuration on its mass. For low-mass objects
belonging to the normal evolutionary sequences, the gravitational mass
is unaffected by the rotation.  For the large rest-mass sequences,
$M\ge 2.4 M_{\odot}$ the gravitational mass increases with the spin
frequency.  Furthermore, the limiting frequency of a (e.g.  $M =
2M_{\odot}$) purely nucleonic star is larger than that for a hybrid
star of the same mass, which is explained by the result that the
nucleonic stars are more compact.
\begin{figure}
\begin{center}
% \includegraphics[height=7cm,width=7cm,angle=0]{nucOmega_vs_TW.eps}
% \vspace{1.cm}
% \includegraphics[height=7cm,width=7cm,angle=0]{Omega_vs_TW.eps}
\includegraphics[keepaspectratio,width=7cm,angle=0]{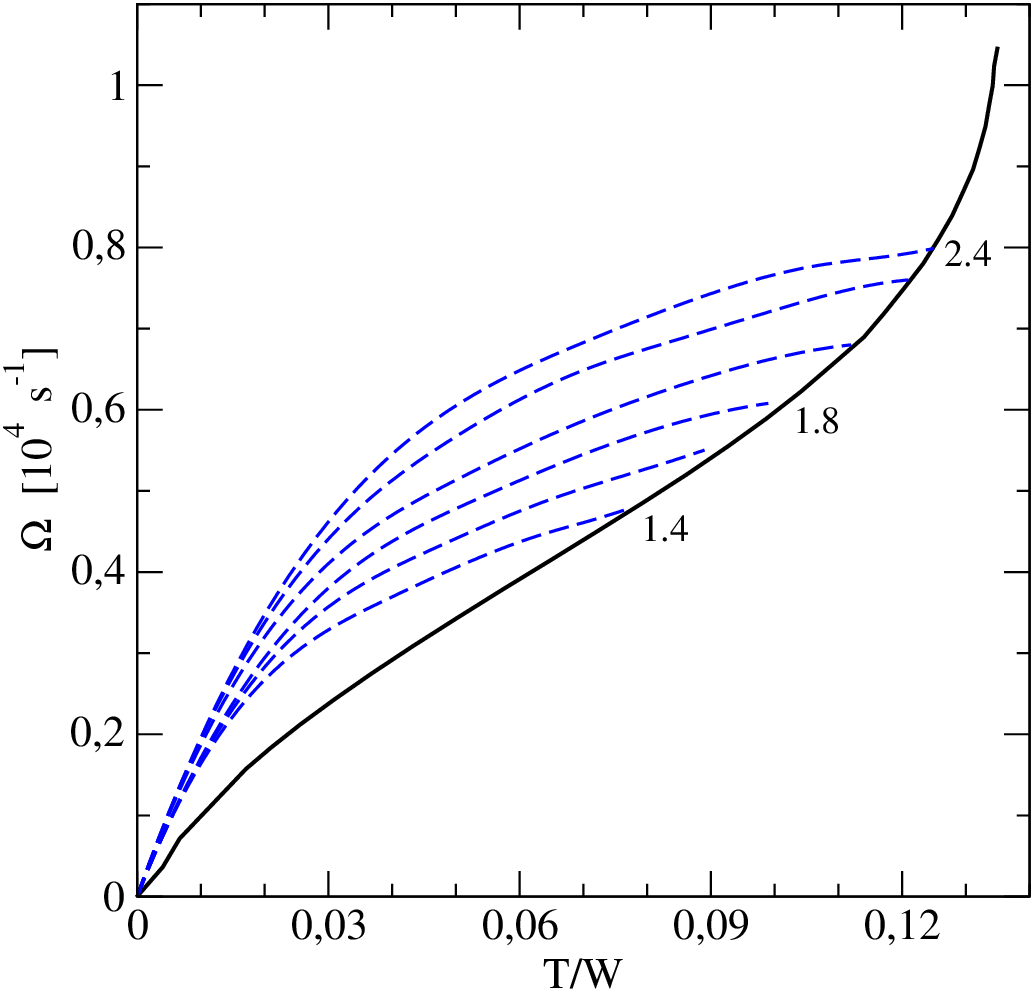}
\vspace{1.cm}
\includegraphics[keepaspectratio,width=7cm,angle=0]{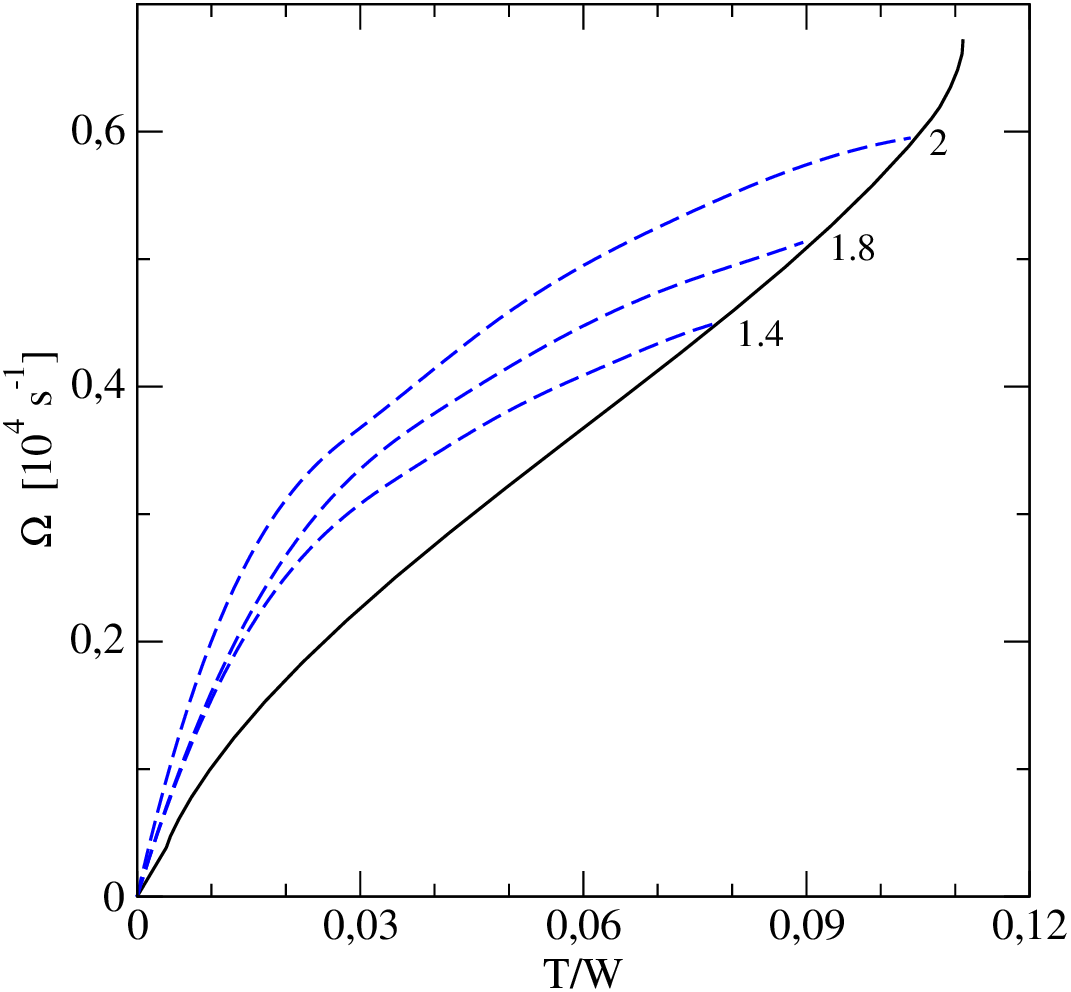}
\end{center}
\caption[]
{Dependence of the spin frequency of configurations on the ratio of
  the kinetic to potential energy $T/W$. 
 The conventions are the same as in Fig.~\ref{fig:Om_J}.
}\label{fig:OM_TW}
\end{figure}
%------------------------------------------------------------------------
\begin{table}
\begin{tabular}{llll}
\hline 
Pulsar   & $M/M_{\odot}$&  $P$(ms)  & Ref.  \\
\hline
J$1740-5340$  & $1.53\pm 0.19$ & 3.65  &   \cite{2003AJ....125.1546K}\\
J$1903+0327$ &  $1.67\pm 0.01$ & 2.15  &   \cite{2008Sci...320.1309C}\\
J$1909-3744$ &  $1.44\pm 0.024$ & 2.95 &   \cite{2003ApJ...599L..99J,2005ApJ...629L.113J}  \\
J$1614-2230$ &  $1.97\pm 0.04$ & 3.15&       \cite{2010Natur.467.1081D}\\
J$1748-2446$ad  &   $\quad -\quad $    & 1.395  &  \cite{2006Sci...311.1901H}\\
\hline
\end{tabular}
\caption{Masses and spin periods of several pulsars taken from the
  references quoted in the fourth column. 
}
\label{tab:PSR}
\end{table}
%------------------------------------------------------------------------
For some pulsars, the mass and the spin frequency are measured. A
representative collection of these objects, which includes the
subgroup of massive and rapidly rotating stars is shown in
Table~\ref{tab:PSR} and in Fig.~\ref{fig:Om_M}. Within this collection,
the most massive pulsar J$1614-2230$ would be, according to our model,
a hybrid star. It is also found to be on a transitional sequence,
since its asymptotically fast-rotating limit corresponds to a purely
nucleonic star. Thus, this pulsar must have
experienced a phase transition to color-superconducting quark phase
 during its evolution. (See the discussion in the next section).

 Figure~\ref{fig:OM_TW} shows the dependence of the angular velocity
 on the ratio of the kinetic energy $T$ to the gravitational binding
 energy $W$.  The latter quantity is especially useful for the
 discussion of the location of the instabilities of the stars toward
 non-axisymmetrical perturbations. For perturbations classified in
 terms of an expansion in spherical functions, the instabilities arise
 to the lowest order for modes that belong to $l=2$ and $m= \pm 2$
 perturbations. In the case of classical uniformly rotating
 incompressible ellipsoids, the onset of secular (i.e., viscosity
 driven) and gravitational wave-emission instabilities occurs at the
 same point along a sequence when $T/W = 0.14$ for $l=2$ modes with $m
 = 2$ and $m=-2$, respectively. For compressible Newtonian and general
 relativistic models, the point of the onset of the gravitational
 instability occurs at lower values of $T/W \simeq 0.08$, which we
 take as a reference for the discussion.  (Viscosity-driven
 instabilities would evolve stars to a non-uniformly rotating
 quasi-equilibrium state, which need not be stationary. The onset of
 secular instabilities in general relativistic stars requires values
 of $T/W$ that are larger than the value 0.08 quoted above for
 gravitational wave-emission instabilities). According to
 Fig.~\ref{fig:OM_TW}, the nucleonic and hybrid models with masses
 $M\ge 1.4 M_{\odot}$ would undergo an instability starting at some
 frequency, whereas lower-mass stars would not. From
 Figs.~\ref{fig:profile1} and \ref{fig:profile2} we conclude that a
 $M\sim 2 M_{\odot}$ mass star, which is predicted to be on the
 transitional sequence, should have been in the unstable region at high
 rotation rates, which corresponds to a purely nucleonic interior.

The evolution of an isolated star would require a slow-down due to the
emission of gravitational waves, which are accompanied by a phase
transition that leads to superconducting quark matter in the core of the star due
to the compression. 

\section{Phase transition to a 
color-superconducting state via spin up or spin down}
\label{sec:PhaseTrans}

Rotationally induced phase transitions to quark matter have been
discussed in the literature and have been claimed to have strong
observational consequences in the timing of isolated pulsars, as well
as accreting X-ray binaries. If the phase transition is weakly first
order (or second order) the EoS of matter may not experience strong
changes as the phase transition sets in. In contrast, the density jump
associated with a strong first-order phase transition may cause sudden
changes in the integral parameters of stars. Previous work
concentrated on the phase transition to normal quark
matter~\citep{1997PhRvL..79.1603G,2000A&A...357..968C,2001ApJ...559L.119G,2002A&A...395..151S,2003NuPhA.715..831G,2006A&A...450..747Z,2009MNRAS.396.2269D,2009MNRAS.392...52A,2009A&A...502..605H,2013arXiv1307.1103W}.
Contrary to this, we argue here that the transition for sufficiently
cold stars occurs to one of the color-superconducting phases of quark
matter.  If the phase of interest is a superfluid, as is the case with
the CFL phase, the onset of superfluidity doubles the dynamical
degrees of freedom of the fluid and  induces a differential
rotation between the normal component, which is coupled to the star on
short dynamical timescales and the superfluid component, which 
couples to the rest of the star via mutual
friction~\citep{2008PhRvL.101x1101M}.  In the case of the 2SC phase,
color magnetism of this phase requires generation of color
magnetic flux tubes in quark matter~\citep{2010JPhG...37g5202A}. This
may alter the magnetic properties of the stars and their spin-down
under electromagnetic radiation. These aspects of the rotationally
induced phase transition deserve separate studies. Here, we restrict
ourselves to a discussion of the internal structure of the stars as
they are spun up or down.

\begin{figure}[t]
\begin{center}
  \includegraphics[keepaspectratio,width=7cm,angle=0]{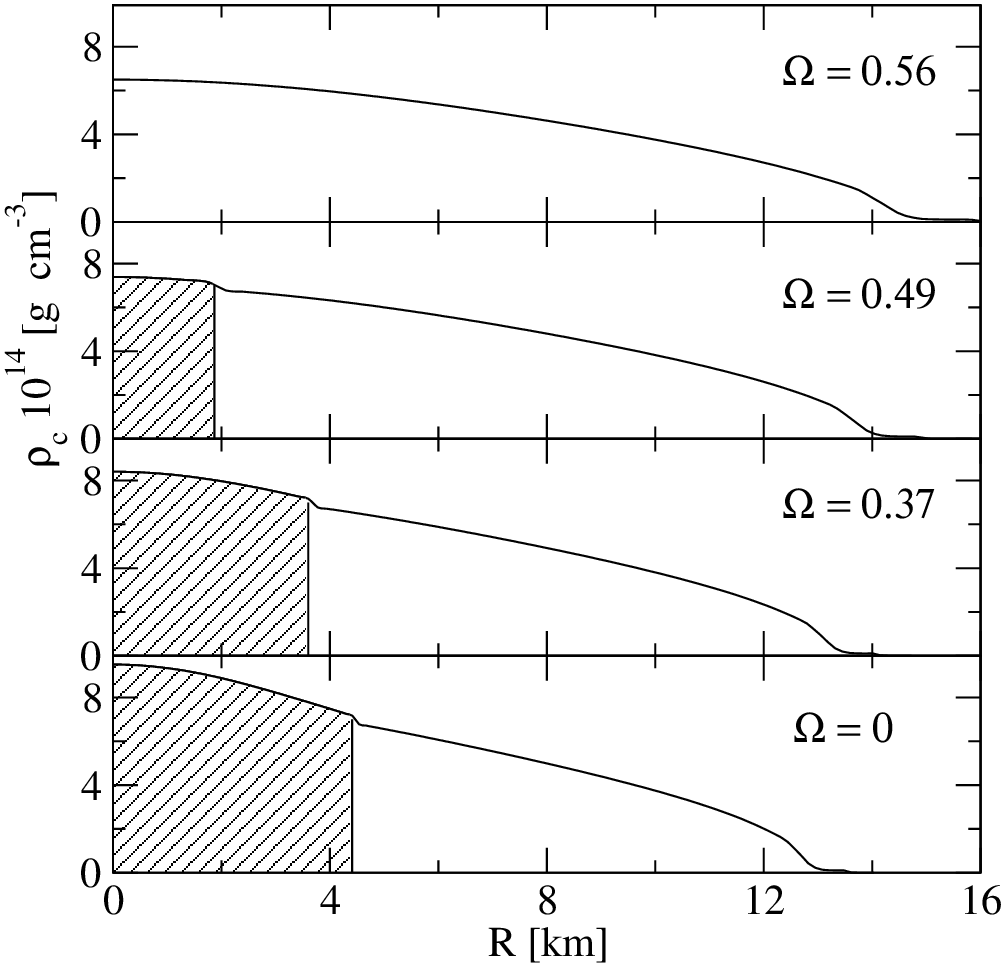}
\end{center}
\caption[]
{ Equatorial density profiles of a rotating compact star of rest mass
  $M_r/M_{\odot} = 2.2$  
at various  rotation frequencies given in units of $10^4$ s$^{-1}$. The shaded
  areas correspond to a quark matter core in the 2SC phase and the empty areas  to
  nucleonic matter.  
}\label{fig:profile1}
\end{figure}

Figure~\ref{fig:profile1} shows the dependence of the equatorial
density profile of a 2.2 $M_{\odot}$ rest-mass star on the internal
radius for four spin frequencies in the range $0\le \Omega \le
5.6\times 10^3$ s$^{-1}$. For the largest spin frequency, it is seen
that the star is purely nucleonic. A reduction of the spin frequency
by about 10$\%$ induces a phase transition to the 2SC phase in the
central region of the star. As expected the stellar radius shrinks
with the spin down. We also note the density jump at the interface
between the 2SC and nucleonic phase as implied by the first-order phase
transition. At asymptotically slow rotation rates, the star represents
a dense 2SC quark phase that extends up to $4$~km and is surrounded by a
nuclear shell enclosed in the region $4\le r \le 13$\ km. The
snapshots of the internal structure of the star displayed in
Fig.~\ref{fig:profile1} would give a true reflection of the internal
structure of the stars, as it spins up or down, if the dynamical 
timescales are much larger than the timescales required for the
nucleation of the color-superconducting phase.

\begin{figure}[t]
\begin{center}
  \includegraphics[keepaspectratio,width=7cm,angle=0]{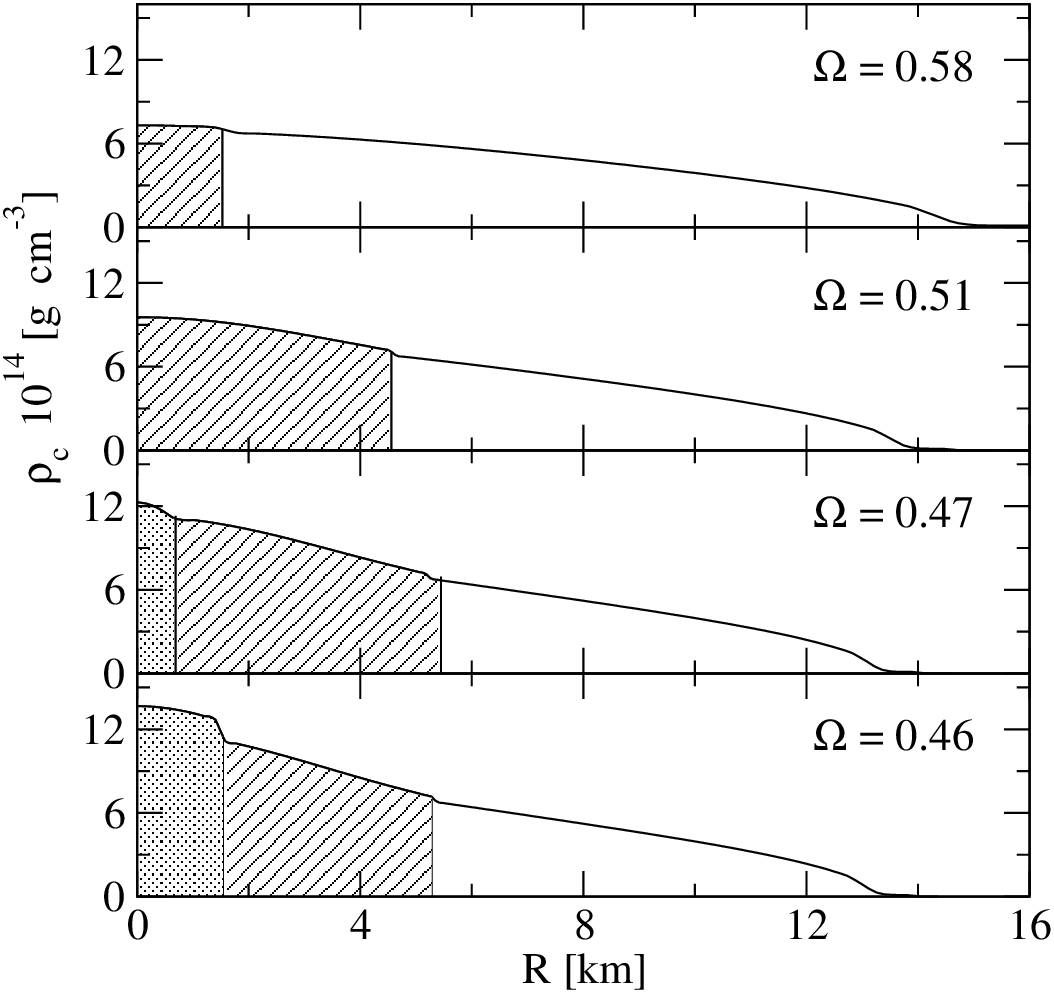}
\end{center}
\caption[] {The same as in Fig.~\ref{fig:profile1} but for a
  configuration with a rest mass $M_r/M_{\odot} = 2.4$.  The hatched
  region corresponds to the 2SC phase and the dotted region to the CFL
  phase.  }\label{fig:profile2}
\end{figure}

Figure~\ref{fig:profile2} displays the same dependence as
Fig.~\ref{fig:profile1} but for a supramassive star with a rest mass
2.4 $M_{\odot}$. In this case, the Keplerian configuration contains a
2SC quark core. As the spin frequency is lowered, the region occupied
by the 2SC phase expands.  A reduction of the spin frequency by 20$\%$
induces a new phase transition to the CFL phase. The phase transition
from the 2SC to the CFL is first order and is accompanied by a jump in
the density at the interface of these two phases.  We recall that
because this configuration is supramassive it does not posses a
static limit.  At about $\Omega = 4.6\times 10^3$ s$^{-1}$, the star
reaches the stability limit. While the CFL phase expanded further, we
see that the 2SC phase started to shrink.  Consequently, there must be
a maximum of the superconducting quark core radius as a function of
the spin frequency along the evolutionary
sequence.  The fate of the supramassive stars is not known:
most likely, they undergo a collapse. Our models can be used as initial
data for simulations of the collapse of hybrid stars with
color-superconducting cores to a black hole.

\section{Conclusions}
\label{sec:conclusion}

We examined the properties of rotating hybrid stars constructed from
a low-density nucleonic EoS and a high-density color-superconducting quark
matter EoS by assuming a first-order phase transition between these
phases at some interface. At high densities, the 2SC phase is replaced
by the CFL phase of superfluid quark matter. We constructed the
normal and supramassive families of purely nuclear and hybrid
configurations. Both families produce massive $M\sim 2M_{\odot}$
compact stars;  the maximum mass of the hybrid stars is close
to this observational limit, while the purely nucleonic family can
sustain masses of order $2.6M_{\odot}$ in the static limit.

The properties of the evolutionary sequences were examined for both
EoSs, where each star loses angular momentum to radiation and slows
down while maintaining its rest mass. The integral parameters of these
sequences and their dependence on the angular rotation frequency agree
with the studies of nucleonic stars. A new feature of these sequences
is the rotationally induced phase transition to a
color-superconducting phase of quark matter, which occurs for
sufficiently massive stars. We note that latent heat, associated with
the first order phase transition, is released as the transition
proceeds from the baryonic matter to 2SC phase and from the 2SC phase
to the CFL phase.

We defined a new subclass of transitional sequences. These are
constant rest-mass sequences that feature a superconducting phase
transition point when the rotation frequency is reduced from the
Keplerian frequency to zero. These sequences are characterized by an
onset and growth of the volume of the superconducting phases as the
star slows down to the static limit. In the case of supramassive
sequences  (for which the static limit is absent)  the radius of the
superconducting quark phase shows a non-monotonic dependence of the
radius (and the volume) of the quark core on the spin frequency.  Our
models of supramassive color-superconducting stars provide the initial
data that can be used to simulate the collapse of an unstable star to
a black hole.  

Because the star contains a superconducting quark core, the collapse
dynamics and potential gravitational wave signal (which are sensitive
to the rotation rate of the star and to its internal composition) may
be affected.  The differential rotation of the superfluid phases and
the existence of density jumps at the phase transition boundaries can
leave a detectable mark on the signal. Moreover, the fast radio bursts
recently observed on cosmological
distances~\citep{2013Sci...341...53T} have been interpreted as a
supramassive compact star that collapses to a black
hole~\citep{2013arXiv1307.1409F}. In this scenario, the
electromagnetic emission is produced by the snapping of the
magnetosphere as the stellar surface collapses. If this suggestion is
correct, a fast radio bursts will provide sufficiently common sources
to explore the presence of a superconducting quark core because the
dynamics of the stellar surface will depend on the internal structure
of the collapsing matter.

\section*{Acknowledgments}

NSA thanks the Frankfurt University for hospitality, where part of
this work was done. AS acknowledges the hospitality of the
Albert-Einstein-Institut, Potsdam. This research was supported by a
collaborative research grant of the Volkswagen Foundation (Hannover,
Germany).

\bibliographystyle{aa}
%\bibliography{rapidrot,pulsar_data,hybrids2,colorsup}{}

\end{document}